\documentclass[aps,pra,twocolumn,groupedaddress,amsmath,amssymb]{revtex4}

\usepackage{epsfig}
\usepackage{amssymb}
\usepackage{amsmath}
\usepackage{graphicx}
\usepackage{amsfonts}
\usepackage{epsfig}
\usepackage{revsymb}
\usepackage{bm}
\usepackage{amsmath}
\usepackage{amssymb}
\usepackage{latexsym}
\usepackage{thumbpdf}
\usepackage{hyperref}

\begin{document}

\title{On the Lifetime of Metastable Metallic Hydrogen}

\author{S.N. Burmistrov}
\author{L.B. Dubovskii}
\affiliation{Kurchatov Institute, 123182
Moscow, Russia}


\begin{abstract}
The molecular phase of hydrogen converts to the atomic metallic phase at
high pressures estimated usually as 300 -- 500~GPa. We analyze the decay of
metallic phase as the pressure is relieved below the transition one. The
metallic state is expected to be in the metastable long-lived state down to
about 10 -- 20~GPa and decays instantly at the lower pressures. The pressure
range of the long-lived metastable state is directly associated with an
impossibility to produce a stable hydrogen molecule immersed into the
electron liquid of high density. For lower pressures, the nucleation of an
electron-free cavity with the energetically favorable hydrogen molecule
inside cannot be suppressed with the low ambient pressure.

\end{abstract}

\maketitle
\section{Introduction}
\par
For the first time, the question on the transition of molecular hydrogen
into metallic phase under pressure was apparently attempted by Wigner and
Huntington \cite{Wi}. Later, in a great amount of papers \cite{Br,Ka} (and
therein) the equation of state for the metallic state as well as the
pressure of the transition into the atomic metallic state are analyzed. In
addition, there has arisen a question whether the lifetime of metastable
metallic phase could be macroscopically large in some pressure range below
the metal-to-molecular phase transition. Here we attempt this problem at
zero temperature.
\par
In paper \cite{Br} the structure of metallic hydrogen is studied in detail
at zero temperature. It is shown in particular that the metallic hydrogen at
zero pressure is energetically stable against the decay into separate atoms
with the binding energy of about 1~eV per atom. As it concerns the decay
into molecules, metallic hydrogen is unstable and the energy of about 2.5~eV
releases with escaping a molecule from the metal surface. The escape of
hydrogen molecule from the surface of metallic hydrogen should occur via
tunneling across a potential barrier. Thus, in general, the lifetime of
metallic phase against this process may prove to be sufficiently large. On
the other hand, this channel can be withdrawn provided the metallic hydrogen
is confined with the corresponding walls.
\par
The formation of molecules is possible not only at the surface of a metal
but also in its bulk. The latter process cannot be eliminated. For the
formation of a molecule inside the bulk of metallic phase, it is necessary
to have a cavity in which the metallic electron density is sufficiently
small so that the molecule would be energetically favorable. The inception
of a cavity is always associated with increasing the total energy in the
system due to extrusion of a metal from the cavity. In order to place a
molecule into the cavity, the latter should have a size of several
interatomic distances. In general, the energy gain resulted from the
formation of a single molecule cannot compensate an increase of the total
energy in the system. In any case this occurs in the pressure region near
the molecular phase-metal phase transition point since the chemical
potentials per atom in the metallic and molecular phases are close to each
other. Thus the nucleation of the molecular phase with the large number of
molecules in the critical nucleus becomes necessary. The large number of
particles in the critical nucleus results inevitably in a drastic reduction
of the nucleation probability of such nuclei because this process is a
tunneling overcoming of a potential barrier and the tunneling probability
depends exponentially on the number of particles.
\par
In the present work we study in what pressure range below the transition
pressure the macroscopic description of the nucleus dynamics is possible and
how this range depends on the approximations chosen. As we will see later,
the lifetime of metallic phase is macroscopically large and practically
infinite so long as the macroscopic consideration is possible.
\par
In the opposite case when the critical nucleus is not large and contains a
few molecules, the lifetime of the metallic phase is small. This can be
estimated as follows. The probability $W_0$ for nucleating the critical
nucleus as large as a single molecule in a specific site of a bulk is always
small
\begin{equation}
W_0\sim\omega _D\exp\bigl(-\alpha\sqrt{m/m_e}\,\bigr). \label{f11}
\end{equation}
The point is that there is a large factor in the exponent, i.e. square root
of a ratio of atom mass $m$ to electron mass $m_e$. Here $\omega _D$ is a
frequency of about Debye frequency in metallic hydrogen ($\omega _D\sim
10^{14}$~s$^{-1}$ and $\alpha$ is the quantity associated with the tunneling
motion of hydrogen atoms in the metallic phase in the course of nucleating a
molecule. We estimate $\alpha\lesssim 1$ since the typical energy barriers
for the motion of nuclei is about 1~eV and distances are of order of 10 --
20~nm. However, the total probability $W_{\nu}$ is large for the inception
of a single nucleus in the bulk containing $\nu\sim 10^{22}$ atoms
\begin{equation}
W_{\nu}\sim\nu W_0. \label{f12}
\end{equation}
This gives a short lifetime for a metallic hydrogen sample with the large
$\nu\sim 10^{22}$ number of atoms.
\par
Note that the same estimate for the small-sized particles of about $\nu\sim
10 ^6$ atoms yields a sufficiently large lifetime. For the process as an
escape of molecules from the surface, the lifetime may also prove to be
large since $\nu\sim 10^{15}$ in this case. In addition, for the evaporation
it is essential not the probability of a single event for the formation of a
molecule but the evaporation rate determined by escaping the large number of
molecules.
\begin{figure}
\includegraphics[scale=0.9]{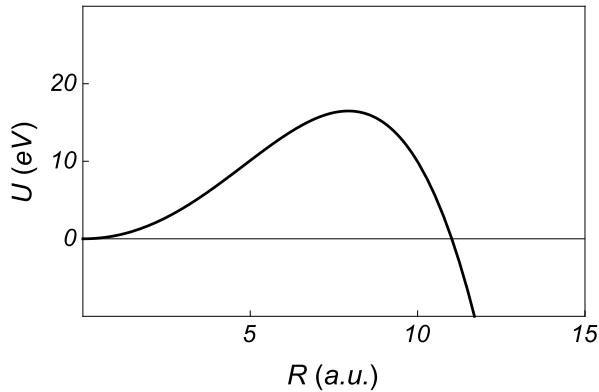}
\caption{The potential energy $U$ versus nucleus radius $R$ at ambient
pressure $P=$~40~GPa. The critical radius is $R_c=$11~a.u. The number of
particles in the critical nucleus is $N_c=$190.} \label{fig1}
\end{figure}
\par
To describe a macroscopic nucleus, we employ the Lifshitz-Kagan approach
\cite{Li}. As a main variable in this approach, we take the density of the
phases, i.e., stable (molecular) and metastable (metallic) ones. The
potential energy of the system as a function of the nucleus radius $R$ has a
typical shape given in Fig.~\ref{fig1}. The growth of potential $U$ at small
radius $R$ is determined by the effective interphase surface tension and
proportional to the radius-squared, i.e. $U(R)\sim R^2$ as $R\rightarrow 0$.
In the case of the junction between the metallic and molecular phases the
effective surface tension is mainly associated with the electron liquid
outflow from the metal and with the decrease of the binding energy of a
molecule in the electron liquid.
\par
The negativity of potential energy $U$ at large radius $R$ is due to
unfavorable difference in energies of metastable metallic and stable
molecular phases, i.e. $U(R)\sim -\Delta\mu\, R^3$ where $\Delta\mu$ is a
difference in the chemical potentials of the both phases. The transition
from metastable state $R=0$ to stable state $R\rightarrow\infty$ occurs via
tunneling under potential barrier (Fig.~\ref{fig1}) due to kinetic energy
$T(R,\dot{R}$ depending on both radius $R(t)$ and growth rate $\dot{R}(t)$
\begin{equation}
T(R,\dot{R})=M(R)\dot{R}^2/2. \label{f13}
\end{equation}
The mass $M(R)$ in the kinetic energy is associated with a difference in the
densities of the metastable and stable phases and results from the outflow
or inflow of the substance during the formation of a nucleus.
\par
Besides the various densities the phase transition can be characterized with
a number of other internal variables independent of density, e.g. spacing
between the nuclei in the course of nucleating a molecule. These internal
variables are characterized with the corresponding potential barriers and
kinetic energies. Below we suppose that the setting in equilibrium in these
variables is the faster process and we take the optimum magnitudes of those
variables. In the next section we elucidate the procedure in detail.
\par
As will be shown below, such macroscopic approach, associated with the
nucleus dynamics governed with the different phase densities, is possible in
a wide range of pressures below the critical one $P_c\sim $300 -- 500~GPa
down to 10 -- 20~GPa. Within this pressure range the critical nuclei have a
large number of particles, resulting in a long-lived stability of metallic
hydrogen. Note that we underrate the pressure range, neglecting a series of
effects which should certainly lead to increasing the lifetime of metallic
phase.

\section{\label{s2} Problem statement}

Let spherical molecular nucleus of radius $R$ be in the metallic hydrogen at
the ambient pressure $P$. The potential energy $U(R)$ of a nucleus can be
written as (\ref{a3})
\begin{gather}
U(R)=4\pi\! \int _0^R\! n(r,R)\biggl(\varepsilon\bigl(n(r,R), R-r\bigr)-\mu
_0(P)\biggr)r^2\, dr \nonumber
\\
+(4\pi /3)PR^3 +4\pi\sigma R^2. \label{f21}
\end{gather}
Here $\mu _0(P)$ is the chemical potential of metallic hydrogen, $n(r,R)$ is
the density of the molecular phase at point $r$ of the nucleus with radius
$R$, $\varepsilon\bigl(n(r,R), R-r\bigr)$ is the energy density of the
molecular phase, and $\sigma (P)$ is the surface tension of the interface.
\par
The magnitude of surface tension $\sigma$ and the behavior of energy density
$\varepsilon\bigl(n(r,R), R-r\bigr)$, as a function of the distance from the
boundary with the metal, are determined with extending the electron liquid
outside the metal into the near-surface region of about Wigner-Seitz radius
$r_s$ in size \cite{La}. For the energy density $\varepsilon$ of molecular
phase, depending on the molecular phase density $n$ and the distance from
the metallic hydrogen boundary, we employ simplest approximation
\begin{equation}
\varepsilon (n, x)=\varepsilon (n)+h(x). \label{f22}
\end{equation}
The first term here corresponds to the energy density of molecular phase for
the given density $n$ in the lack of the metal electron density. The second
term $h(x)$ implies that the energy of a molecule beside the metal boundary
differs significantly from energy 4.7~eV in vacuum taken from the energy of
two separate atoms due to dipping a molecule into electron liquid of a
metal. The term $h(x)$ can be represented as an external potential affecting
the molecule as a result of extending the electron liquid outside the metal
into the boundary region of about $r_s$ in size. Thus, potential $h(x)$  as
well as surface tension $\sigma$ depend on the Wigner-Seitz radius $r_s$ or,
correspondingly, on the pressure $P$ inside the metal.
\par
In addition, these both quantities, $\sigma$ and $h(x)$, depend  on the
nucleus size $R$ as well. We will neglect this dependence since we are
interested in the macroscopic $R\gg r_s$ nuclei and the dependence of
$\sigma$ and $h(x)$ on radius $R$ becomes insignificant as $R\gtrsim r_s$.
\par
Relation (\ref{f22}) corresponds to the gas approximation in the density of
molecular phase, meaning a possibility to neglect dependence of $h(x)$ on
density $n$. The approximation can be used while the density of molecular
phase at the boundary is much smaller than the density of the adjacent metal
phase. The point is that potential $h(x)$ is governed with the outflow of
electron liquid from the metal, which is almost independent of the strongly
localized electron density at the molecule \cite{Hj,No1}. In the
near-surface region, where the magnitude $h(x)$ is large, the density of
molecular phase takes the smaller value compared with that at the nucleus
center since such density distribution corresponds to the minimum of
potential energy $U(R)$. At low $P\lesssim 100$~GPa pressures there appears
a gap of about $r_s$ between the molecular and metallic phases. Inside the
gap the density of hydrogen atoms vanishes. For the higher pressures, it is
impossible to assert that the density of molecular phase beside the nucleus
boundary is much smaller than the density of metallic phase. However, even
near the molecular phase-to-metallic phase transition point $P_c$ the
density of molecular phase differs from that of metallic phase by a factor
of 2. In the next section we discuss function $h(x)$ in detail since this
quantity governs mainly the nucleation probability.
\par
Due to the same reason we will neglect the dependence of surface tension
$\sigma$ on the molecular phase density at the nucleus boundary, i.e. we put
the surface tension equal to its magnitude for the vacuum-metal boundary.
For the dependence of the energy density of molecular phase $\varepsilon
(n)$ upon $n$, we apply the local approximation $\varepsilon =\varepsilon
\bigl( n(r,R)\bigr)$ since the involvement of, e.g., density gradient in
$\varepsilon (n)$, corresponds to considering the quantities in such scale
which we neglect in describing the potential $h(x)$. This is also associated
with the smallness of density gradient in the nucleus due to large nucleus
radius $R\gg r_s$.
\par
The expression (\ref{f21}) for the potential energy of a nucleus assumes
that the internal energy of the system depends on the densities of phases
alone. On the whole, this implies the liquid-like description of the both
phases. The involvement that the both phase are the crystalline  ones
increases the potential energy. The liquid-like description of the system
means neglecting the shear energy compared with the energy of the bulk
compressibility which is completely taken into account in Eq.~(\ref{f21}).
(See Appendix \ref{A1}.) Such neglect in metallic hydrogen is always
justified since the shear modulus is small as compared with the bulk modulus
\cite{Br,Ka}. In molecular hydrogen under low $P\lesssim 10$~GPa pressures
the both energies, bulk compressibility energy and shear energy, are small
compared with the energy of formation of molecules and are inessential in
the expression for potential energy. For larger $P\gtrsim 10$~GPa pressures,
there occurs the same situation as in metallic hydrogen. The shear energy is
small compared with the bulk compressibility energy and can be neglected as
before.
\par
We will suppose that the nucleus grows slowly, namely, the growth rate
$\dot{R}$ of nucleus boundary is much less than the sound velocity $s$
 $$
\dot{R}\ll s .
 $$
In the case of such quasistationary nucleus growth there is a sufficient
time to set the mechanical equilibrium in the bulk of the both phases. An
existence of mechanical equilibrium in the metastable metallic phase is
taken in Eq.~(\ref{f21}) into account since the metallic density $n_0$ as
well as pressure $P$ are assumed to be constant in the derivation of
Eq.~(\ref{f21}). As it concerns a nucleus, the equilibrium over the nucleus
corresponds to constancy of the total chemical potential
\begin{equation}
\mu\bigl(n(r,R)\bigr)+h(R-r)=C. \label{f24}
\end{equation}
Here $C$ is a constant to be determined with the conditions at the nucleus
boundary.
\par
Equation (\ref{f24}) can be obtained with varying the potential energy
$U(R)$ over nucleus density $n(r,R)$ in (\ref{f21}). On the other hand, this
equation is a result of the hydrodynamical Euler equation \cite{La2}
\begin{equation}
\frac{\partial\bm{v}}{\partial t} +(\bm{v}\nabla )\bm{v}=-\frac{1}{n}\nabla
P -\nabla h. \label{f25}
\end{equation}
Putting $\bm{v}\rightarrow 0$, we arrive at Eq.~(\ref{f24}) since $\nabla
P/n=\nabla (\varepsilon +n\partial\varepsilon /\partial n)$ resulted from
relation $P=n^2\partial\varepsilon /\partial n$. Besides Eq.~(\ref{f24}) the
minimum in energy $U$ can also be associated with $n\equiv 0$ in some region
of a nucleus due to condition $n\geqslant 0$.
\par
Equation (\ref{f24}) and $n =0$ allow us to determine the density
distribution $n(r,R)$ in the nucleus bulk. Constant $C$ in (\ref{f24}) can
be found using the condition of mechanical equilibrium between the nucleus
and the metastable phase. The condition of mechanical equilibrium between
the phases is obtained with varying the potential energy $U$ in the nucleus
radius  $R$ under invariant total number $N$ of particles in the nucleus
\begin{gather}
\bigl(\partial U/\partial R\bigr)_N=0, \label{f26}
\\
N=4\pi\int _0^Rn(r,R)r^2\, dr. \label{f27}
\end{gather}
\par
Equation~(\ref{f26}) means the following. As the nucleus radius varies, the
system keeps the state of the minimum potential energy but has no sufficient
time to perform the transition of particles from one phase to the other.
Emphasize that this condition holds due to assumption about the
quasistationary nucleus growth when the growth rate is small compared with
the sound velocity $s$. The latter velocity characterizes the rate of
setting the mechanical equilibrium. Equation (\ref{f26}) determines the
relation between constant $C$ (\ref{f24}) and pressure $P$ in the metallic
phase (Appendix~\ref{B1})
\begin{gather}
P=-\frac{2\sigma (P)}{R}-\frac{1}{R^2}\int _0^Rn(r)h'(R-r)r^2\, dr \nonumber
\\
+n(R)\bigl[C-\varepsilon\bigl(n(R)\bigr)-h(0)\bigr]. \label{f28}
\end{gather}
Here $n(r) =n(r,R)$ and $n(R)=n(R,R)$.
\par
Equations (\ref{f24}), (\ref{f28}) and $n=0$ determine the density
distribution $n(r,R)$ in the nucleus at the ambient pressure $P$. If we
substitute the above distribution into Eqs.~(\ref{f21}) and (\ref{f27}), we
obtain the dependence of energy $U$ and particle number $N$ upon nucleus
radius $R$. The typical behavior of energy $U$ as a function of radius $R$
is given in Fig.~\ref{fig1}. The point $R_c$ at which $U(R_c)=0$ corresponds
to the critical nucleus radius.
\par The role of temperature in the quantum transition from the metastable
phase with $R=0$ to the state of overcritical nucleus with $R>R_c$ is
replaced with the kinetic energy in variables $R$ and $\dot{R}$ resulting
from the different densities of the phases and outflow of a matter from the
nucleus \cite{Li}. The kinetic energy of a nucleus is given by
\begin{gather}
T=T_1+T_2, \label{f29}
\\
T_1=\frac{m}{2}4\pi\int _0^Rn(r,R)v^2(r) r^2\, dr, \nonumber
\\
T_2=\frac{m}{2}4\pi\int _R^{\infty}n_0(P)v_0^2(r) r^2\, dr. \nonumber
\end{gather}
Here $m$ is the hydrogen atom mass, kinetic energy $T_1$ is associated with
the motion of particles in the nucleus, and $T_2$ is due to the motion of
the mass in the metallic phase. Velocity $v(r)$ inside the nucleus is
determined by the continuity equation
\begin{equation}\label{f210}
\frac{\partial n(r,R)}{\partial t}+\frac{1}{r^2}\frac{\partial}{\partial
r}\bigl(r^2v(r)n(r,R)\bigr)=0.
\end{equation}
Remind that density $n$ depends on time $t$ via variable $R=R(t)$ alone and
$r$ is a running coordinate. Thus, we have
\begin{equation}\label{f210a}
\frac{\partial n}{\partial t}=\frac{\partial n}{\partial R}\dot{R} .
\end{equation}
The solution of the above two equations for velocity $v(r)$ at a given
distribution $n(r,R)$ can be written as
\begin{equation}\label{f211}
v(r)=\frac{\dot{R}}{r^2n(r,R)}\int _0^r\frac{\partial n(r',R)}{\partial
R}r^{\prime 2}\, dr'.
\end{equation}
The substitution of Eq.~(\ref{f211}) into (\ref{f25}) shows that the
left-hand side of Eq.~(\ref{f25}) is small on the scale of a ratio
$\dot{R}/s\ll 1$. So, we have a quasistationary growth of a nucleus and, at
first, we can find distribution $n(r,R)$ obeying (\ref{f24}). Then, we
substitute the distribution obtained into Eq.~(\ref{f211}) to find the
velocity distribution $v(r)$ and avoid the combined solution of the Euler
equation (\ref{f25}) and continuity equation (\ref{f210}). Substituting
(\ref{f211}) into the equation for $T_1$ yields
 $$
T_1=\frac{m}{2}4\pi\dot{R}^2\int_0^R\frac{dr}{r^2n(r,R)}\biggl(\int
_0^r\frac{\partial n(r',R)}{\partial R}r^{\prime 2}\, dr\biggr)^2.
 $$
\par
The velocity distribution $v_0(r)$ in the metastable phase obeys the
continuity equation (\ref{f210}) as well. However, unlike (\ref{f210a}) the
term with the time derivative of the density vanishes since the density in
the metastable phase is constant for all $R$ and is determined with the
ambient pressure $P$. This entails the following behavior \cite{Li}
\begin{equation}\label{f212}
v_0(r)=A\dot{R}R^2/r^2.
\end{equation}
The dimensionless factor $A$ can be found using the condition of conserving
the total number of particles in the system
\begin{equation}\label{f213}
dN/dt=4\pi R^2n_0(P)\bigl(\dot{R}-v_0(R)\bigr).
\end{equation}
The left-hand side equals the rate of varying the particle number $N$ in a
nucleus and the right-hand side does the incoming flow of particles from the
metastable phase. Since the time dependence in (\ref{f27}) enters via
variable $R$, we arrive at
\begin{equation}\label{f214}
\frac{dN}{dt}=4\pi R^2n(R)\dot{R}+4\pi\dot{R}\int _0^R\frac{\partial
n(r,R)}{\partial R}r^2\, dr.
\end{equation}
Using Eqs.~(\ref{f212}) -- (\ref{f214}), we obtain for $A$
\begin{equation}
A=1-\frac{n(R)}{n_0(P}-\frac{1}{R^2n_0(P)}\int _0^R\frac{\partial
n(r,R)}{\partial R}r^2\, dr.
\end{equation}
Putting Eq.~(\ref{f212}) into the expression for $T_2$ and calculating the
integral in $r$, we represent kinetic energy $T$ in (\ref{f29}) as
\begin{gather}\label{f29a}
T=M(R)\dot{R}^2/2,
\\
M(R)= 4\pi m\biggl[n_0(P)R^3\biggl(1-\frac{n(R)}{n_0(P)} \nonumber
\\
-\frac{1}{R^2n_0(P)} \int _0^R\frac{\partial n(r,R)}{\partial R}r^2\,
dr\biggr)^2 \nonumber
\\
+\int_0^R\frac{dr}{r^2n(r,R)}\biggl(\int_0^r\frac{\partial n(r',
R)}{\partial R}r^{\prime 2}\, dr'\biggr)^2\biggr]. \nonumber
\end{gather}
This expression differs from the corresponding one in \cite{Li} because the
compressibility of the stable phase is taken into account. In our case the
involvement of compressibility is essential since the density of molecular
hydrogen varies by a factor of 10 from zero pressure to the 100~GPa pressure
region.
\par
The next analysis of nucleation kinetics is based on the hamiltonian with
the potential energy (\ref{f21}) and kinetic energy (\ref{f29a}). The
density distribution in the nucleus bulk is governed with Eqs.~(\ref{f24})
and $n=0$ and related with the ambient pressure via Eq.~(\ref{f28}).
\par
Deriving the hamiltonian, we have assumed that the state of the system is
completely determined with the densities of the phases and the other
physical quantities are adjusted adiabatically and unambiguously to the
magnitudes of the densities. As an example of such quantities, we can
mention the spacing between two atoms in the hydrogen molecule, symmetry of
the crystalline lattice in the both phases, and electron density tracing
adiabatically the nuclear motion. The adiabatical adjustment of these
quantities supposes the slow and quasistationary nucleus growth when the
setting of all processes in equilibrium occurs faster than the nucleus
growth, i.e., transition of particles from the metallic phase to the
molecular one. In particular, this implies the mechanical equilibrium
between phases (\ref{f26}). The adiabatical relaxation of these parameters
means the neglect of contributions of these parameters both to the potential
and to the kinetic energies. Such optimization underrates the lifetime of
the metastable phase. Setting the equilibrium in the these parameters occurs
at about sound velocity or faster as, for example, in the case of adiabatic
relaxation of electrons to the motion of nuclei. Hence, we suppose the
smallness of nucleus growth rate $\dot{R}$ compared with the sound velocity
$s$. Adiabaticity and relaxation of all parameters are equivalent to the
fact that the frequency of oscillations, associated with the underbarrier
motion, is much smaller than all other frequencies in the system. The
frequency $\omega _b$, determining the nucleus underbarrier evolution,
reduces as the nucleus radius $R_c$ grows
\begin{equation}\label{f216}
\omega _b\sim\biggl(\frac{U_{max}}{M(R_c)R_c^2}\biggl)^{1/2}\sim\omega
_D\biggl(\frac{r_s}{R_c}\biggr)^{3/2}.
\end{equation}
Here $U_{max}\sim\sigma R_c^2$ and $M(R_c)\sim mn_0R_c^3$ where $m$ is an
atom mass and $\omega _D$ is of the order of the Debye frequency. For the
nucleus of large radius $R_c$, frequency $\omega _b$ is small, entailing a
correctness of the quasistationary application. In the case when the size of
the critical nucleus is about several interatomic distances it is necessary
to take the lack of quasistationary approximation into account.

\section{\label{s3}Discussion of the parameters in the hamiltonian}

The potential energy $U(R)$ (\ref{f21}) and kinetic energy $T(R,\dot{R})$
(\ref{f29a}) are governed with the following parameters such as:
(\textit{i}) chemical potential of metallic phase $\mu _0(P)$ depending on
the pressure, (\textit{ii}) energy $\varepsilon (n)$ and chemical potential
$\mu (n)$ depending on the density, (\textit{iii}) energy variation $h(x)$
of a molecule beside the metal surface, and (\textit{iv}) surface tension
$\sigma (P)$ of a metal. Below we consider the consistent description of the
above quantities.
\begin{figure}
\includegraphics[scale=0.9]{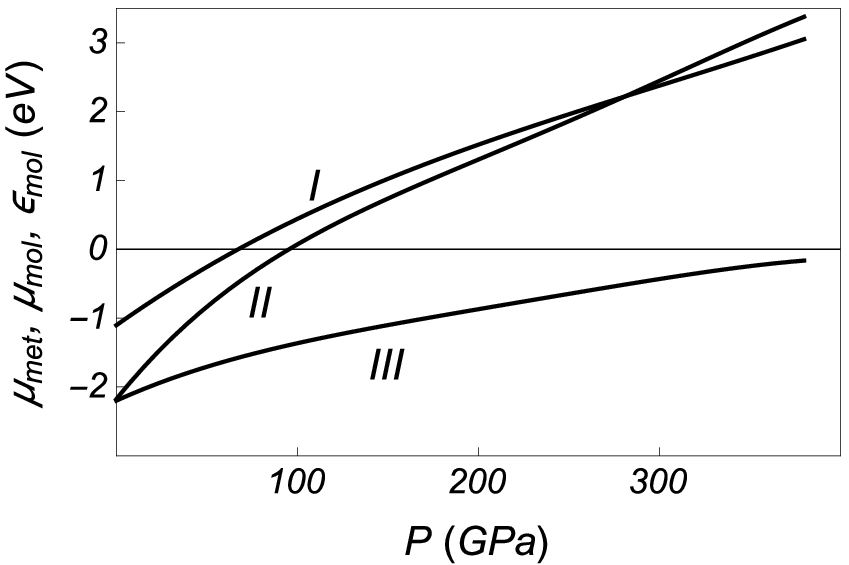}
\caption{The chemical potentials of metallic (\textit{I}) and molecular
(\textit{II}) phases as a function of pressure. Curve \textit{III} is the
energy of molecular phase. \label{fig2a}}
\includegraphics[scale=0.9]{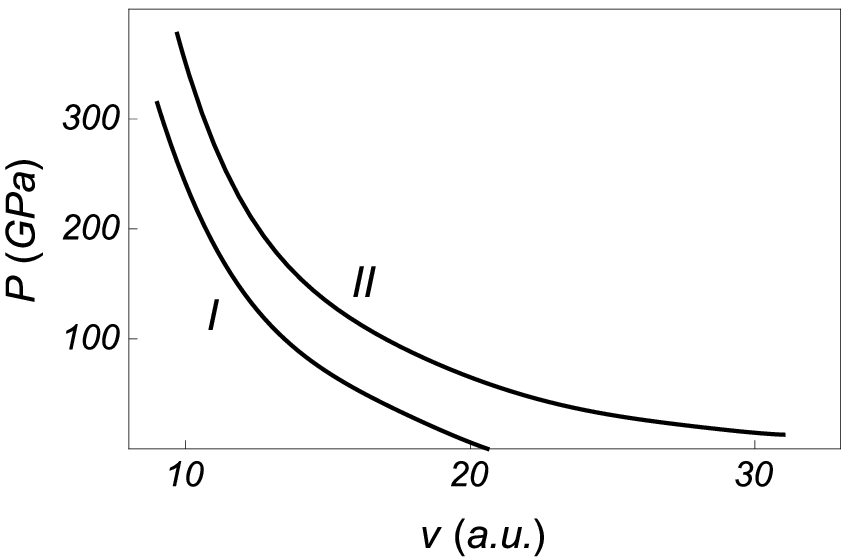}
\caption{The equations of state for metallic (I) and molecular (II)
phases.\label{fig2b}}
\end{figure}
\par
The behavior of energy and chemical potentials as a function of pressure are
given for the molecular and metallic phases in Fig.~\ref{fig2a}.
Figure~\ref{fig2b} plots the equations of state for the both phases.
\par
We take the results for the metallic phase after \cite{Ka}. For the
molecular phases, we use the results from the same work \cite{Ka} and also
after \cite{An}. Note that the functions given for metallic hydrogen are
obtained with high accuracy at high $P\gtrsim$~100~GPa pressures and,
correspondingly, at small $r_s<1.45$. In zero pressure region of $r_s\approx
1.7$ the accuracy of the quantities calculated is smaller and the error for
the equation of state within this range may be estimated as $\pm$5~GPa.
\par
On the contrary, the equation of state for the molecular phase is well-known
in the low pressure region as $P\lesssim$10~GPa. This is associated with the
existence of precise hydrostatic measurements and with the relatively exact
description since it is sufficient mainly to consider the pair interactions
alone. For higher $P\gtrsim 10$~GPa pressures, the consideration of pair
interactions alone becomes insufficient \cite{An} and the theoretical
description of the equation of state has a worse accuracy. The same is
referred to the experiments in this region. An uncertainty in the data for
the equation of state results in a dispersion of the phase transition
pressure $P_c$ \cite{Ka}. However, for our purposes such dispersion has no
principal meaning since the pressure region where the metallic phase exists
practically the infinite time is wide and extends down to pressures of about
10~GPa. An uncertainty of the equations of states for the phases shifts the
pressure boundary by $\pm$5~GPa as the transition pressure of about 300 --
500~GPa. Note that the boundary of the pressure region where the metastable
metal exists in the long-lived state is drastic since the lifetime depends
on the pressure via large exponent $\beta$. This entails that the small
variations of the exponent results in a strong variation of the exponential
expression.
\par
\begin{figure}
\includegraphics[scale=0.9]{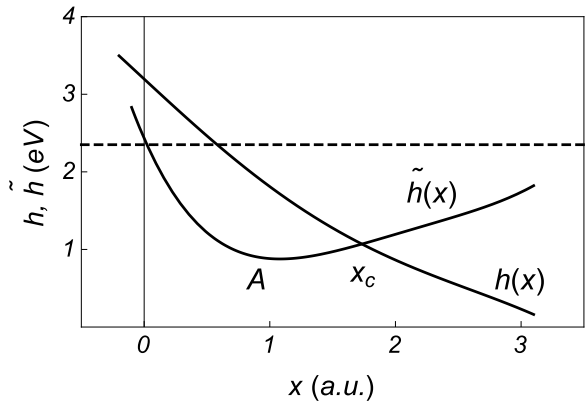}
\caption{The functions $h(x)$ and $\tilde{h}(x)$ as a function of distance
from the metal surface. The metallic phase is approximated with the jellium
model of $r_s=1.7$. The dash line corresponds to the energy of a hydrogen
atom far from the metal surface.} \label{fig3}
\end{figure}

The main parameter, which determines the region of the long-lived metastable
state, is the energy of a molecule $h(x)$ varying beside the metal surface
due to spreading the electron density outside the metal. The magnitude
$h(x)$ can be obtained with the direct calculation of the behavior of the
energy of a molecule as a function of the distance taken from the metal
surface. Such calculation is performed in \cite{Hj} using the model of
jellium. Figure~\ref{fig3} shows the dependence $h(x)$ extrapolated from the
data after \cite{Hj} to $r_s=1.7$ corresponding to that of metallic hydrogen
at zero pressure \cite{Br}. In this figure we also show the energy of a
separate hydrogen atom beside the metal jellium surface with $r_s=1.7$
denoted as $\tilde{h}(x)$.
\par
Minimum $A$ at the curve $\tilde{h}(x)$ (Fig.~\ref{fig3}) corresponds to the
energy position of the chemical potential for atoms in metallic hydrogen. At
this point there occurs a chemical sorption of hydrogen molecule at the
metallic hydrogen surface. This means a possibility of adding new layer to
the metallic hydrogen surface. Minimum $A$ of curve $h(x)$ lies higher than
the magnitude $h(x)$ at $x\rightarrow\infty$. Otherwise, there occurs an
associative chemical sorption at the surface, i.e. chemical sorption with
releasing the energy at the transition of a particle from the surface to the
infinity. If atoms are located far from the surface as compared with the
intersection point $x_c$ of curves $h(x)$ and $\tilde{h}(x)$
(Fig.~\ref{fig3}), the existence of a molecule becomes possible. For the
distances closer than $x_c$, the separate atoms are more energetically
favorable.
\par
\begin{figure}
\includegraphics[scale=0.9]{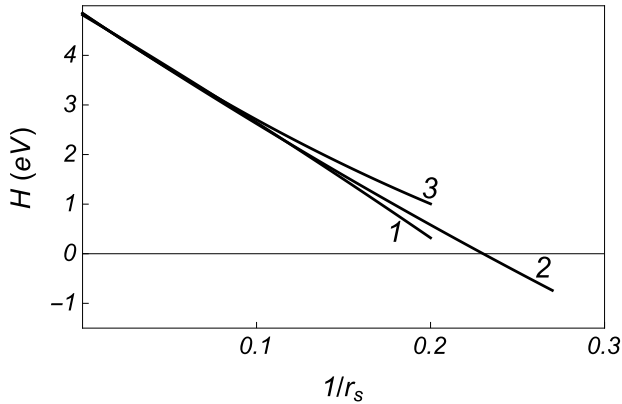}
\caption{The plot of binding energy $H$ for a molecule immersed into an
electron liquid.} \label{fig4}
\end{figure}

The energy $h(x)$ of a molecule in the field of electron liquid can be
subdivided into the energy of atoms in the molecule and the binding energy
of a molecule
\begin{equation}\label{f31}
h(x)=\frac{1}{2}\bigl[\tilde{h}(x-R_0/2)+\tilde{h}(x+R_0/2)\bigr]+H.
\end{equation}
Here $R_0$ is the equilibrium distance between the nuclei in the molecule.
The binding energy $H$ of a molecule~\cite{No1} proves to be slightly
affected with the orientation of a molecule. Thus, the local approximation
governed by the electron liquid density $\rho (x)$ is well adequate, i.e.,
\begin{equation}\label{f31a}
H=h\bigl(\rho (x)\bigr).
\end{equation}
The behavior $H(\rho )$, as a function of $r_s$, is shown in Fig.~\ref{fig4}
and $\rho^{-1}=4\pi r_s^3/3$. The different curves in Fig.~\ref{fig4}
correspond to various methods of calculating the function $H$.
\par
Curves 1, 2 and 3 are obtained from the data \cite{Hj}  on the behavior of a
hydrogen molecule beside the surface of metallic jellium with various
$r_{s0}$ of a metal, namely, 2.07, 2.65 and 4. Subscript 0 in $r_{s0}$
differs $r_{s0}$ of metallic substrate from $r_s$ determining  the magnitude
of electron density with the aid of $\rho ^{-1}=4\pi r_s^3/3$. One can see
that the curves, obtained with the different ways, are very close to each
other and one may say about the universal behavior.
\par
In order to derive the behavior of energy of a molecule as a function of the
distance from the metal surface if one knows behavior $H(\rho )$, it is
sufficient to apply the behavior $\rho (x)$ beside the surface \cite{La}.
The latter problem is one-dimensional and, therefore, is simpler.
\par
One can see from Fig.~\ref{fig4} that the binding energy of a molecule
vanishes at the electron density $r_s=4.6$. For smaller $r_s$, molecule is
energetically unfavorable and dissociates into atoms.
\par
Note that the behavior of $h(x)$ and $\tilde{h}(x)$ in Fig.~\ref{fig3}
correlates well with the data \cite{Br} on metallic hydrogen. The asymptotic
behavior $h(x)$ for $x\rightarrow\infty$ is determined with the binding
energy of hydrogen molecule and the position of minimum $A$ is governed by
the binding energy of an atom in metallic hydrogen. The accuracy of
coincidence between the minimum at curve $\tilde{h}(x)$ and the position of
the chemical potential for atoms in metallic hydrogen is determined by the
neglect of the coupling between atoms in the surface layer. This
approximation is analogous to neglecting the dependence $h(x)$ on the
density of molecular phase $n$.
\par
While obtaining the plots given in Figs.~\ref{fig3} and \ref{fig4}, the
genuine metal with the discrete structure is replaced with the jellium
model. This approximation is well justified at the large distances from the
metal. In essence, if $x>r_s$, the discreteness of the crystal lattice
becomes insignificant. In addition, function $h(x)$ at such large distances
is of most interest. The point is that in the range of relatively low
pressures of about 10~GPa, there appears a spacing between the metal and
molecular phases in which the density of molecular phase vanishes. Thus, for
such pressures an uncertainty due to inaccurate determination of function
$h(x)$ at small distances   is negligible. For higher pressures, the
lifetime grows more and an additional specifying $h(x)$ in this range
becomes inessential. The growth of $h(x)$ at small $x<r_s$ distances in the
metal with the discrete lattice is reduced as compared with the jellium
model since the discrete ions are located from the molecule farther on than
for the smoothed background of the jellium. The discrete ions attract
electrons stronger and, therefore, electron liquid spreads at smaller
distance from the surface as compared with the jellium model. This results
in the slower enhancement of function $h(x)$ in the metal with the discrete
lattice than that in the jellium model. Below we take this fact into account
and vary function $h(x)$ at small distances in order to clarify its effect
on the lifetime of the metallic phase.
\par
While calculating $h(x)$ in \cite{Hj}, the axis of a molecule is assumed to
be normal to the metal surface. Provided one neglects the effect of the
molecule orientation on the distribution of electron liquid beside the metal
surface and takes into account that the binding energy $H$ of a molecule
depends only on the magnitude of the electron liquid density and is
independent of the molecule orientation, one can obtain information about
$h(x)$ for an arbitrary orientation using the data on $h(x)$ for the
normal-to-surface orientation of a molecule. For example, in the case when
the molecule axis is parallel to the surface the following relation is valid
\begin{equation}
h_{\parallel}(x)=\tilde{h}(x)+H(x).
\end{equation}
\par
In the calculation \cite{Hj} of function $h(x)$ the parameter $R_0$,
distance between the nuclei in the molecule, keeps unvaried as the molecule
approaches the metal surface. This approximation is well justified due to
slight dependence of the binding energy on $R_0$.
\par
To conclude the discussion of functions $h(x)$ and $\tilde{h}(x)$, we note a
few aspects. First, the binding energy of a molecule is positive starting
from the electron densities with $r_s=4.6$. For lower $r_s$, molecule is
energetically unfavorable. This specific magnitude $r_s$ is three times as
larger than $r_s$ of metallic hydrogen. Correspondingly, the electron liquid
density at the center of cavity, in which the molecule could be placed,
should be at least $\approx 30$ times as smaller if compared with the
electron density of metallic hydrogen. Thus, to nucleate a single molecule
inside the metallic phase, it is necessary to produce a large cavity  with
the radius of a few $r_s$. This fact correlates with the assumption in the
previous sections that the nucleation of the molecular phase requires the
outflow of a matter in the metallic phase and that the radius of the
critical nucleus should significantly exceed $r_s$.
\par
The next point to be mentioned is that the hydrogen atom escaping from the
metallic phase and traveling from point A (Fig.~\ref{fig3}) along curves
$h(x)$ and $\tilde{h}(x)$ should overcome an energy barrier. In our model
(Sec.~\ref{s2}) we neglect a possibility  for reflection of hydrogen atom
from the energy barrier in the course of quantum tunneling. This implies
that we underrate the lifetime of the metastable metallic phase.
\par
Here we emphasize also that the real crossover between the curves $h(x)$ and
$\tilde{h}(x)$ is smooth-like since an additional parameter $R_0$, distance
between the nuclei in the molecule, varies with the distance from the metal
surface. In principle, there are possible two situations for the transition
from the molecular to metallic phase.
\par
First, $R_0$ varies smoothly from the typical spacing between the nuclei in
a molecule to that in the metallic phase. Second, variation $R_0$ is a
jump-like one at the phase interface. Below, as in the previous section, we
imply that distance $R_0$ follows adiabatically the density. Then a single
distinction between these two cases is the following. As distance $R_0$
varies continuously, function $h(x)$ behaves more smoothly in the narrow
transient region between the phases. So, from this viewpoint it is useful to
vary $h(x)$ at distance of about $r_s$ in the transient region between  the
phases. This will be done below with the analysis of the data.
\par
\begin{figure}
\includegraphics[scale=0.9]{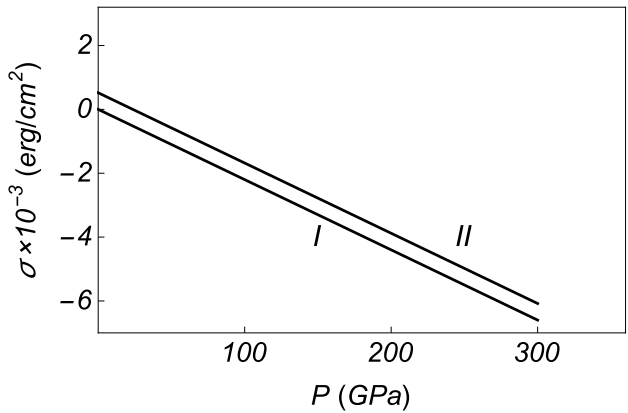}
\caption{The surface tension-pressure dependence.} \label{fig5}
\end{figure}

Finally, we discuss the surface tension $\sigma (P)$ of metallic hydrogen
for the vacuum-metal boundary. The calculation of the surface tension at the
vacuum-metal boundary and comparison with the experimental data has been
treated in a large number of works \cite{La}, \cite{Hi} and \cite{Bu}. In
all these papers the consideration is based on the Hohenberg-Kohn-Sham
density functional in which the kinetic, exchange and correlation energies
of nonuniform electron gas are described as a functional of electron density
$\rho (\bm{r})$. The surface tension of a metal results from the
redistribution of electrons and ions beside the metal surface as compared
with the bulk distribution. We here employ the simplest version of
Ref.~\cite{Bu} when the electron distribution is assumed to be homogeneous
in the metal bulk. In this case the surface tension $\sigma$ as a function
of pressure is given with curve~\textit{I} in Fig.~\ref{fig5}. The curve is
well fitted with the relation
\begin{equation}\label{f35}
\sigma _I(P)=-23P \;\;\;\text{in erg/cm}^2.
\end{equation}
Here pressure $P$ is given in GPa. This approximation neglects a series of
contributions to the surface tension. The main contribution neglected is
that the density of exchange and correlation energies in nonuniform electron
gas has a nonlocal relation with the electron liquid density. Provided this
contribution is taken into account as a simple gradient correction, we
obtain the surface tension-pressure plot as a curve~\textit{II} in
Fig.~\ref{fig5}. The curve is well described with
\begin{equation}\label{f35a}
\sigma _{II}(P)=-22P +520 \;\;\;\text{in erg/cm}^2.
\end{equation}
Here we do not discuss the finer effects associated, e.g., with a shift of
the edge ion planes beside the metal surface \cite{Hi} or with nonuniform
distribution of electrons in the metal bulk \cite{Bu} since these
contributions are smaller than the term resulted from the gradient
exchange-correlation energy. In addition, these corrections become
insignificant due to uncertainty  of the equation of state in the zero
pressure range. Emphasize that the shift of the equation of state with about
5~GPa results in varying the surface tension by about 100 erg/cm$^2$ within
the zero pressure range. In what follows, we use mainly
expression~(\ref{f35}) for the surface tension since this expression results
in the stricter condition for the lifetime of the metastable metallic phase.
\par
Note that the final result for the lifetime of metastable metallic hydrogen
is not noticeably sensitive whether one takes Eq.~(\ref{f35}) or
Eq.~(\ref{f35a}) for the surface tension. The point is that the function
$h(x)$ itself results in the effective surface tension which exceeds surface
tension $\sigma (P)$ by a few times for the moderate pressures of about
10~GPa. For the larger pressures, the effect of surface tension $\sigma (P)$
is weaker due to effect of the bulk term $4\pi\rho R^3/3$.

\section{\label{s4} Quantum nucleation of the molecular phase. The discussion of
results}
\par
Like \cite{Li}, we perform the semiclassical analysis of the tunneling
transition between the phases. The classical Lagrangian of the system reads
 $$
L(R,\dot{R})=M(R)\dot{R}^2/2 -U(R).
 $$
Here effective mass $M(R)$ is determined with Eq.~(\ref{f29a}) and energy
$U(R)$ is given by Eq.~(\ref{f21}). In Appendix~\ref{C1} the derivative
$\partial n(r,R)/\partial R$ in the equation for mass $M(R)$ is transformed
to the ambient pressure-fixed expression. The hamiltonian corresponding to
the above Lagrangian reads
\begin{equation}\label{f41a}
H=p_R^2/2M(R)+U(R).
\end{equation}
\par
Within our approximation the dynamic description of the system during phase
transition is governed with the single principle variable $R=R(t)$ and
corresponding momentum $p_R$. The initial state of the system is a
metastable state. Provided a possibility of the tunneling transition is
ignored, the ground state in potential $U(R)$ for radius $R$ close to
zero~(\ref{fig1}) can be estimated with using the uncertainty principle as
 $$
p_R\cdot R_{\text{typ}}\sim \hbar .
 $$
Since radius $R$ is not large, one can approximate potential $U(R)$ as
 $$
U(R)=4\pi R^2_{\text{typ}}.
 $$
Denoting the ground state energy as $E_0=\hbar\omega_0$, we have
 $$
\omega _0\sim 4\pi\sigma R_{\text{typ}}^2/\hbar .
 $$
Due to
 $$
\frac{p_R^2}{2M(R_{\text{typ}})}\sim 4\pi\sigma R_{\text{typ}}^2
 $$
and using $M(R)=4\pi mn_0(P)R_{\text{typ}}^3$, we obtain
 $$
R_{\text{typ}}\sim\bigl[\hbar ^2/\bigl(32\pi ^2\sigma
mn_0(P)\bigr)\bigr]^{1/7}
 $$
and
\begin{equation}\label{f46}
\omega _0\sim (16\pi ^3)^{1/7}\sigma ^{5/7}\hbar
^{-3/7}\bigl(mn_0(P)\bigr)^{-2/7}.
\end{equation}
The estimate for $\omega _0$ coincides with the semiclassical expression in
\cite{Li}. Note that unlike $\omega _b$~(\ref{f216}), frequency $\omega _0$
is independent of critical radius $R_c$. The point is that frequency $\omega
_0$ is associated with the heterophase quantum fluctuations in the
homogeneous metastable phase and is insensitive to critical radius $R_c$. In
the semiclassical approximation the probability for the quantum transition
from level $\hbar\omega_0$ to nucleation of the critical nucleus is given by
\begin{equation}\label{f47}
W=\nu\omega _0\exp\bigl(-\beta \bigr), \;\;\; \beta =
\frac{2}{\hbar}\int_0^{R_c}|p_R|\, dR.
\end{equation}
Equation~(\ref{f46}) yields
\begin{equation}\label{f46a}
\omega _0\sim\omega _D(m/m_e)^{3/14}\sim 10^{14}s^{-1}.
\end{equation}
Here $m/m_e$ is a ratio of proton mass to electron one and $\nu$ is the
number of virtual nucleation centers of new phase. The latter is of the
order of the number of particles in the system. Thus, the preexponential
factor coincides with Eqs.~(\ref{f11}) and (\ref{f12}) and is about
10$^{36}$ particle/s.
\begin{figure}
\includegraphics[scale=0.92]{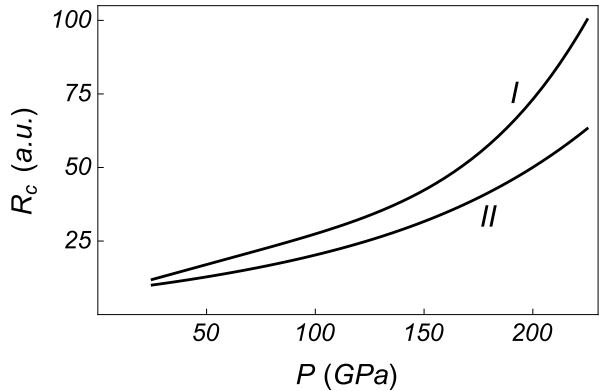}
\caption{The critical radius $R_c$ versus ambient pressure.} \label{fig6}
\end{figure}
\par
Momentum $p_R$ is determined semiclassically for the state of energy $E$
close to zero, $\hbar\omega _0\ll U_{\text{max}}$, as
 $$
|p_R|=\sqrt{2M(R)U(R)}.
 $$
The integration in~(\ref{f47}) is performed over the positive region of
potential energy $U(R)>0$. To have a macroscopically long-lived state of
metastable metallic phase, the exponent $\beta$ in~(\ref{f47}) should be
large and not smaller than 80 - 100. For the smaller exponents, the large
preexponential factor in~(\ref{f47}) compensates the effect of the exponent,
resulting in the large decay probability.
\begin{figure}
\includegraphics[scale=0.9]{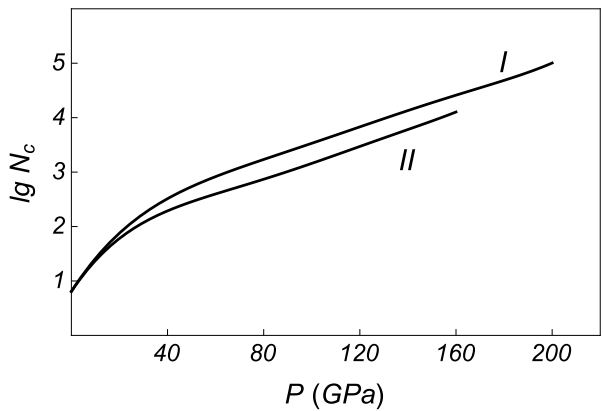}
\caption{ The number of particles $N_c$ in the critical nucleus as a
function of ambient pressure (logarithmic scale).} \label{fig7}
\end{figure}
\par
The main parameter determining the exponent is the critical nucleus radius
$R_c$ or critical number of particles $N_c$. The $R_c$-$P$ dependence is
plotted in Fig.~\ref{fig6}. Curve \textit{I} corresponds to choice $h(x)$
obtained by extrapolating the data to metallic hydrogen $r_s$ at zero
pressure $r_s=1.7$ (Fig.~\ref{fig3}) and to the surface tension determined
with the curve~\textit{I} in Fig.~\ref{fig5}. Curve \textit{II} in
Fig.~\ref{fig6} is obtained with the same surface tension but with the
function $h(x)$ changed for $x<1.5$~a.u. Function $h(x)$ is truncated and
put equal to $h(x=1.5\, a.u.)$ (Fig.~\ref{fig3}). Such variation $h(x)$
remains the critical radius $R_c$ unchanged (Sec.~\ref{s3}) at low pressures
of about $\lesssim$10~GPa when the region for a long-lived existence of the
metastable metallic hydrogen is determined. Thus for determining the
boundaries of stability of stable existence of metallic state, the behavior
of function $h(x)$ is essential only at large distances from the surface.
The behavior at large distances, as noted in Sec.~\ref{s3}, is well-known.
For the higher pressures when critical radius $R_c$ is already large, the
truncation mentioned results in some reduction of critical radius $R_c$.
\begin{figure}
\includegraphics[scale=0.89]{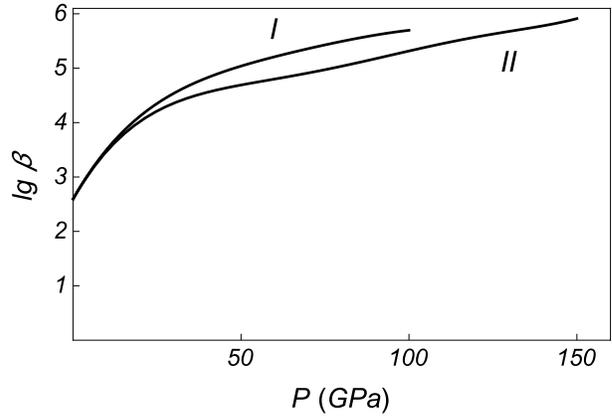}
\caption{The exponent $\beta$ (\ref{f47}) versus ambient pressure
(logarithmic scale).} \label{fig8}
\end{figure}
\par
In Figs.~\ref{fig7} and \ref{fig8} we give the plot of the number of
particles $N_c$ in the critical nucleus and the plot of the exponent $\beta$
in Eq.~(\ref{f47}) for different functions $h(x)$. From Figs.~\ref{fig7} and
\ref{fig8} we can obtain the relation between $\beta$ and $N_c$. It proves
to be that in a wide pressure range this relation can approximately be
described with the linear law $\beta =\alpha N_c$, $\alpha$ being 200 as
$N_c\lesssim 100$ and $\alpha$ being 120 as $N_c\lesssim 10^5$.
\begin{figure}
\includegraphics[scale=0.54]{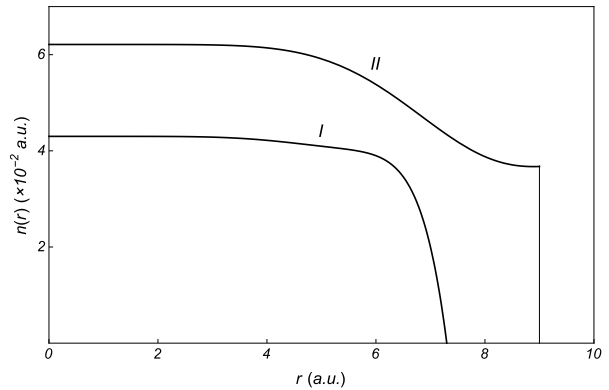}
\caption{The density distribution in the nucleus of radius $R=9$ at
pressures 20 GPa (curve \textit{I}) and 75 GPa (curve \textit{II}).}
\label{fig9}
\end{figure}
\par
\begin{figure}
\includegraphics[scale=0.58]{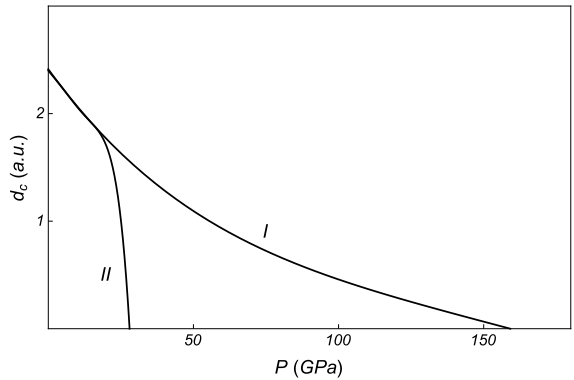}
\caption{The spacing $d_c=d(R_c)$ for the critical nucleus as a function of
pressure for various functions $h(x)$ (curves \textit{I} and \textit{II}).}
\label{fig10a}
\end{figure}

In Fig.~\ref{fig9} we plot the typical distribution for the molecular phase
density $n(r, R)$ inside the nucleus at various pressures. It is seen that,
for the relatively low pressures,  there is a spacing $d$ where the density
of the molecular phase vanishes. For the pressures larger than 75~GPa, the
density of molecular phase does not vanish everywhere. In Fig.~\ref{fig10a}
we show the dependence of spacing $d_c=d(R_c)$ as a function of pressure $P$
for the critical nucleus.
\par
\begin{figure}
\includegraphics[scale=0.9]{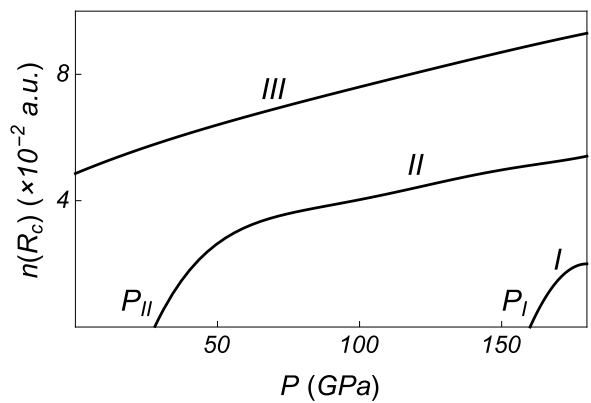}
\caption{The dependence of the minimum magnitude of density $n(R_c)$ of
molecular phase in the critical nucleus for various functions $h(x)$ (curves
\textit{I} and \textit{II}). The density of metallic phase versus pressure
is shown with curve \textit{III}. For $P<P_{I}, n(R_c)=0$ (curve \textit{I})
and for $P<P_{II}, n(R_c)=0$ (curve \textit{II}).} \label{fig10b}
\end{figure}

The plot in Fig.~\ref{fig10b} demonstrates the dependence of the minimum
magnitude of density $n(R_c)$ of molecular phase in the critical nucleus.
Within the whole pressure range the density at the boundary of molecular
phase is significantly smaller as compared with the density of metallic
phase (curve \textit{III} in Fig.~\ref{fig10b}). The latter, as is noted in
Sec.~\ref{s2}, makes it possible to neglect the dependence of function
$h(x)$ on the density of molecular phase.
\par
As the data show, the variation of function $h(x)$ within the reasonable
limits affects insignificantly the main result, i.e. long-lived stability of
existing the metastable metallic phase within the wide pressure range below
the transition point $P_c\sim$300~GPa down to pressure $\sim$10~GPa.  The
variation of surface tension $\sigma (P)$ affects the results to slight
degree. Emphasize that we have used the surface tension (Fig.~\ref{fig5},
curve \textit{I}), resulting in the minimum magnitude of the lifetime.

\section{Summary}

We have analyzed stability of the hydrogen metallic state against nucleation
of the stable molecular phase below the transition pressure $P_{c}\sim$ 300
-- 500 GPa. The nucleation dynamics is governed by the tunneling of a
critical molecular nucleus through a potential barrier in the
low-temperature region and by thermal activation mechanism at high
temperatures. In a wide $0.1 P_c\lesssim P\leqslant P_c$ pressure region
below the phase transition pressure $P_c$ the critical nucleus of the
molecular phase contains a large number of particles and has,
correspondingly, a large critical radius as compared with the interatomic
spacing. The main reason for the large critical nucleus lies in the
impossibility to form a bound state of two hydrogen atoms under high
extrinsic electron density of the metallic phase $r_s\sim 1.7$. This entails
the necessity to produce a cavity inside the metallic phase with the low
electron density in the center insomuch that the formation of molecules
would become energetically favorable. The nucleation dynamics of molecular
nuclei at  both low and high temperatures can be described within the
framework of the macroscopic approach. Within the mentioned $0.1 P_c\lesssim
P\leqslant P_c$ pressure region the lifetime of the metallic hydrogen phase
is macroscopically large and the metallic state is practically stable, i.e.
long-lived.
\par
In the low pressure region $P\lesssim 0.1P_c$ the inception of a cavity in
the metallic state cannot be suppressed with the applied external pressure
$P$ and the critical nucleus amounts to a few particles or less as the
external pressure $P$ vanishes. Thus, we expect the opposite behavior with
too small lifetime of the metastable metallic state, resulting in
practically instant decay of the metallic phase.

\appendix
\section{\label{A1}}

Let $V_0$ and $n_0$ be volume and density of the metastable phase. After the
nucleation of stable phase of volume $V'™$ and density $n'$ the volume of
metastable phase becomes $V$ and density does $n$. The energy of nucleus can
be written as
\begin{gather*}
U=\int _V\varepsilon _0(n)n\, d^3r +\int _{V'}\varepsilon '(n')n'\, d^3r
\nonumber
\\
+ \int _{V'}\sigma\, dS' -\int _{V_0}\varepsilon _0(n_0)n_0\, d^3r.
\end{gather*}
Here $\sigma$ is the surface tension, $\varepsilon _0(n)$ and $\varepsilon
'(n')$ are the energy density of the metastable and stable phase,
respectively. Expanding $\varepsilon _0(n)$ in small $(n-n_0)$ as
 $$
\varepsilon _0(n)=\varepsilon _0(n_0)+(n-n_0)P_0/n_0^2
 $$
and substituting it into the first equation, we have
\begin{gather*}
U=-\int _{V+V'}\varepsilon _0(n_0)n_0\, d^3r+\int _{V'}\varepsilon '(n')n'\,
d^3r +
\\
\!\int _V\varepsilon _0(n_0)\bigl[n_0+(n-n_0)\bigr]d^3r
+\frac{P_0}{n_0}\!\int _V(n-n_0)d^3r+\sigma dS'
\\
=\int _{V'}\varepsilon '(n')n'\, d^3r-\int _{V'}\varepsilon _0(n_0)n_0\,
d^3r
\\
+\int _V\biggl(\varepsilon
_0(n_0)+\frac{P_0}{n_0}\biggr)(n-n_0)d^3r+\sigma\int _{V'}dS'.
\end{gather*}
Since
\begin{gather*}
\int _V(n-n_0)d^3r=\int _Vn\, d^3r -\int _{V+V'} n_0\, d^3r+\int _{V'}n_0\,
d^3r
\\
=\int _{V'}n_0\,d^3r - \int _{V'}n'\, d^3r ,
\end{gather*}
we have finally
\begin{equation}
U=\!\int _{V'}\bigl[\varepsilon '(n') -\mu _0(P_0)\bigr]n'\, d^3r +P_0\!\int
_{V'}d^3r +\sigma\!\int_{V'}\! dS' \label{a3}
\end{equation}
taking $\varepsilon _0(n_0)+P_0/n_0=\mu _0(P_0)$ into account.

\section{\label{B1}}

It follows from Eq.~(\ref{f24}) that density $n(r,R)$ of molecular phase
depends on $C$ as a parameter
 $$
n=n(r, R, C).
 $$
Therefore the potential energy $U$ of a nucleus and the number of particles
$N$ depend also on $C$ as a parameter
\begin{gather}
U=U(R,C),\label{b2}
\\
N=N(R,C).\label{b3}
\end{gather}
Using the last relation, parameter $C$ can be expressed via the total number
of particles
\begin{equation}\label{b4}
C=C(R,N).
\end{equation}
Substituting Eq.~(\ref{b4}) into (\ref{b2}), we find $U$ as a function of
$R$ and $N$
 $$
U=U\bigl(R, C(R,N)\bigr).
 $$
\par
The condition of mechanical equilibrium~(\ref{f26}) reads
 $$
\left(\frac{\partial U}{\partial R}\right)_N=0.
 $$
Then we have
\begin{gather*}
4\pi R^2\biggl(n(R)\bigl[\varepsilon\bigl(n(R)\bigr)+h(0)\bigr]-\mu
_0(P)n(R)+P\biggr)
\\
+8\pi\sigma (P)R +\int_0^R n(r)h'(R-r)4\pi r^2\, dr
\\
+\int_0^R\biggl(\mu\bigl(n(r)\bigr)+h(R-r)-\mu
_0(P)\biggr)
\\
\times\biggl[\biggl(\frac{\partial n}{\partial R}\biggr)_C +\frac{\partial
n}{\partial C}\biggl(\frac{\partial C}{\partial R}\biggr)_N\biggr]4\pi r^2\,
dr=0
\end{gather*}
where $n(r)\equiv n(r,R)$ and $n(R)=n(R,R)$. In the last integral the
expression in the parentheses is constant due to~(\ref{f24}) and can be put
in the front of integral. The magnitude of the remaining integral can be
found with differentiating Eq.~(\ref{b3}) in $R$
 $$
\int_0^R\biggl[\biggl(\frac{\partial n}{\partial R}\biggr)_C+\frac{\partial
n}{\partial C}\biggl(\frac{\partial C}{\partial R}\biggr)_N\biggr]4\pi r^2\,
dr +4\pi R^2 n(R)=0.
 $$
Then we obtain Eq.~(\ref{f28}) for pressure $P$ which can be rewritten in
the convenient form for numerics
\begin{gather*}
P=-\frac{2\sigma (P)}{R}-\frac{1}{R^2}\int
_0^R\bigl[r^2n(r)-R^2n(R)\bigr]h'(R-r)\,
dr
\\
+n(R)\bigl[C-\varepsilon\bigl(n(R)\bigr)-h(R)\bigr].
\end{gather*}

\section{\label{C1}}

In the kinetic energy~(\ref{f29a}) the nucleus mass depends on derivative
$\partial n(r,R)/\partial R$, ambient pressure $P$ being fixed. In
Eq.~(\ref{f24}) the density is directly expressed in terms of $C$ related to
pressure $P$. So, it is necessary to transform $\partial n/\partial R$ from
one variable to another
\begin{multline}\label{c1}
\left(\frac{\partial n}{\partial R}\right)_P=\frac{\partial (n,P)}{\partial
(R,P)}= \frac{\partial (n,P)}{\partial (R,C)}/\frac{\partial (P,R)}{\partial
(C,R)}
\\
=\biggl[\biggl(\frac{\partial n}{\partial R}\biggr)_C\biggl(\frac{\partial
P}{\partial C}\biggr)_R -\biggl(\frac{\partial n}{\partial
C}\biggr)_R\biggl(\frac{\partial P}{\partial
R}\biggr)_C\biggr]/\biggl(\frac{\partial P}{\partial C}\biggr)_R
\\
=\biggl(\frac{\partial n}{\partial R}\biggr)_C - \biggl(\frac{\partial
n}{\partial C}\biggr)_R \biggl(\frac{\partial P}{\partial
R}\biggr)_C/\biggl(\frac{\partial P}{\partial C}\biggr)_R .
\end{multline}
Differentiating Eq.~(\ref{f24}) in $R$ under fixed $C$ and then in $C$ under
fixed $R$, we have
\begin{gather*}
\left(\frac{\partial n(r,R)}{\partial R}\right)_C=-\frac{h'(R-r)}{\mu
'(n)}\, ,
\\
\left(\frac{\partial n(r,R)}{\partial C}\right)_R=\frac{1}{\mu '(n)}\, .
\end{gather*}
Derivatives $\bigl(\partial P/\partial R\bigr)_C$ and $\bigl(\partial
P/\partial C\bigr)_R$ are found with differentiating Eq.~(\ref{f28}). Though
$\sigma$ and $h$ depend on $P$, the derivatives of $\sigma$ and $h$ in $P$
do not enter the ratio $\bigl(\partial P/\partial R\bigr)_C/\bigl(\partial
P/\partial C\bigr)_R$. One can see this directly using the cumbersome
calculation
\begin{widetext}
\begin{gather*}
\frac{\bigl(\partial P/\partial R\bigr)_C}{\bigl(\partial P/\partial
C\bigr)_R}=\biggl\{2\sigma (P)-R^2n(R)h'(R) +2Rn(R)\bigl[h(R)-h(0)\bigr]
-\int _0^R\bigl[r^2n(r)-R^2n(R)\bigr]h''(R-r)\, dr
\\
+\frac{2}{R}\int _0^R\bigl[r^2n(r)-R^2n(R)\bigr] h'(R-r)\, dr +\int
_0^R\frac{h^{\prime 2}(R-r)}{\mu '(n)}r^2\,
dr\biggr\}\times\biggl(R^2n(R)-\int _0^R\frac{h'(R-r)}{\mu
'\bigl(n(r,R)\bigr)}r^2\, dr\biggr)^{-1}.
\end{gather*}
\end{widetext}
Substituting the above three relations for the derivatives into
Eq.~(\ref{c1}), we obtain the relation for $\partial n/\partial R$ which
should be employed for calculating $M(R)$~(\ref{f29a}) under fixed pressure
$P$.

\end{document}